# ZnO@C/PVDF Electrospun Membrane as Piezoelectric Nanogenerator for Wearable Applications

*Anshika Bagla[1], Kaliyan Hembram[2], François Rault[3], Fabien Salaün[3*], Subramanian Sundarrajan [4,5], Seeram Ramakrishna [4*], Supratim Mitra[1*]*

[1]*Department of Physical Sciences, Banasthali Vidyapith, Rajasthan 304022, India*

[2] *Center for Nanomaterials, International Advanced Research Center for Powder Metallurgy & New Materials (ARCI), Hyderabad 500005, India*

[3] *Univ. Lille, ENSAIT, ULR 2461 - GEMTEX - Génie et Matériaux Textiles, F-59000 Lille, France*

[4]*Center for Nanofibers & Nanotechnology, Department of Mechanical Engineering, National University of Singapore, Singapore 117574, Singapore*

[5]*Department of Prosthodontics, Saveetha Dental College and Hospitals, Saveetha Institute of Medical & Technical Sciences, Saveetha University, Chennai 600077, India*

***Correspondence(s):*** supratimmitra@banasthali.in, seeram@nus.edu.sg, fabien.salaun@ensait.fr

**Abstract**

The rapid growth of wearable technology demands sustainable, flexible, and lightweight energy sources for various applications ranging from health monitoring to electronic textiles. Although wearable devices based on the piezoelectric effect are widespread, achieving simultaneous breathability, waterproof, and enhanced piezoelectric performance remains challenging. Herein, this study aims to develop a piezoelectric nanogenerator (PENG) using ZnO nanofillers in two morphologies (nanoparticles and nanorods), with a carbon coating (ZnO@C) core-cell structure to enhance piezoelectric performance. Electrospinning technique was employed to fabricate a lightweight, breathable, and water-resistant ZnO@C/PVDF membrane, enabling in situ electrical poling and mechanical stretching to enhance electroactive β-phase formation and thus improve piezoelectric performance. A maximum power density of 384.83 µW/cm$^3$ was obtained at $R_L$ = 10$^4$ kΩ, with a maximum V$_{out}$ = 19.9 V for ZnO@C nanorod-incorporated PVDF samples. The results demonstrate that ZnO@C nanorods exhibit superior voltage output due to their larger surface-to-volume ratio, leading to enhanced interaction with PVDF chains compared to nanoparticles. The fabricated membrane showed promising results with a water vapor transmission rate (WVTR) of ~0.5 kg/m²/day, indicating excellent breathability, and a water contact angle of ~116°, demonstrating significant waterproofness. These findings highlight the potential of the ZnO@C/PVDF electrospun membrane as an effective piezoelectric nanogenerator and energy harvester for wearable applications.

# 1. Introduction

The development of wearable technology has increased many folds in recent years due to substantial demand for smart, portable, and flexible devices. These devices cater to a wide range of applications, including health monitoring, human-machine interfaces, consumer electronics, and defence communication, particularly in the form of electronic textiles [1-5]. However, a major hurdle in the widespread commercialization of wearable devices lies in powering them with sustainable, flexible, and lightweight energy sources. Conventional batteries, with their limited capacity, lifetime, and bulkiness, do not meet the requirements. Consequently, there is a growing interest in energy harvesting technologies that can harness energies form their surrounding environment, such as human motion, mechanical vibrations, converting it into electric energy to provide a sustainable power source [6-14]. Ideally, such energy harvesting solution should be polymer-based to ensure flexibility and lightweight properties.

Among various energy harvesting technologies, piezoelectric energy harvesters are more appealing and have been widely explored due to its excellent capacity to convert ambient mechanical energy to electrical energy and vise versa, making them ideal for self-powered wearable sensors and energy harvesters [15, 16]. Piezoelectric polymers, such as poly (vinylidene fluoride) (PVDF) and its co-polymers, are excellent candidates for energy harvesters owing to their flexibility, lightweight nature, high piezoelectric coefficient, and physical and chemical stability [17-19]. During synthesis, PVDF predominantly crystallizes into the non-electroactive α-phase, which must be converted into electroactive β-phase, along with a minor fraction of γ-phase, to exhibit piezoelectric properties. Additionally, high electric field poling and mechanical stretching are commonly employed [20]. Enhancing piezoelectric properties often involves the use of various piezoceramic fillers such as $BaTiO_3$ (BT), $NaNbO_3$, lead zirconate titanate (PZT), and ZnO [20-23].

Among these, ZnO is particularly promising for wearable application due to its biocompatibility, chemical stability, skin friendliness, antibacterial properties, and large polymer-filler interaction area in composite [24-28]. Mansuri *et al.* utilized ZnO-nanoparticles within a PVDF matrix for energy harvesting, achieving a maximum output power of 32 nW/cm$^2$ [29]. Similarly, Kim *et al.* employed ZnO nanorods in PVDF for wearable applications, obtaining an output voltage of 6.36 V [30]. Alam *et al.* introduced ZnO nanoparticles into PVDF for energy harvesting applications [31] and reported an output voltage of 4.8 V, while Islam *et al.* used ZnO-PVDF composite to fabricate PENGs, yielding an output voltage of 4.2 V [32]. Despite these advancements, the piezoelectric performance of ZnO-polymer composite-based PENGs have been restricted by microstructural variability of the ZnO ceramics.

To improve performance of PENGs, researchers have extensively explored conductive fillers, including carbon-based materials such as graphene, and carbon nanotube (CNT), often combined with ceramic fillers as binary additives. This strategy aims to achieve a synergistic improvement by enhancing the electroactive phase content, providing an excellent electrical conductivity to facilitate more charge transfer between piezo-filler and electrodes, and augmenting the dielectric properties of the composite. For instance, Bhaduri *et al.* demonstrated the incorporation on Mn-doped $BaTiO_3$ (BT) and CNT to develop a self-powered fluoride ion detector, emphasizing the advantage of hybrid fillers in enhancing the piezoelectric output and functionality [33]. Bouhamed *et al.* successfully employed a collaborative network using MWCNT into $BaTiO_3$/PDMS composite to achieve enhanced performance through enhanced conductivity that promotes charge transfer [34]. Similarly, Sun *et al.* highlighted the role of CNTs in enhancing charge transfer and device performance in a flexible ZnO nanoparticle-based nanogenerator [35], while Park et al. demonstrated enhanced piezoelectric output using $BaTiO_3$ nanoparticles combined with graphitic carbon [36]. Li *et al.* also demonstrated the superior piezoelectric output of hierarchical ZnO@CF/PVDF composites for self-powered meteorological sensors, attributing the enhancement to the conductive carbon fiber [37]. Among these strategies, core shell carbon coating on ceramic nanofillers have demonstrated exceptional effectiveness in addressing performance bottlenecks in PENGs. Such coating offers superior electrical properties by reducing internal resistance and enhancing interfacial polarization of the ceramic particles by providing a nano-dimensional bridges between piezo-filler and electrodes. For example, Li *et al.* reported that carbon coated $BaTiO_3$-based core-shell $BaTiO_3$/P(VDF-TrFE) composites significantly improved performance in flexible piezoelectric nanogenerators due to better charge transfer, enhancing interfacial polarizations and filler dispersion [38]. Furthermore, Zhou *et al.* emphasized the role of interface-modulated ceramic nanoparticles (BT@C, PZT@C, KNN@C) in PDMS-based composites for efficient energy harvesting [39]. Advanced hybrid designs, such as MXene-based fillers embedded in polymer matrices [40] underscore the versatility and efficacy of core-shell designs.

Based on these advancements, this study introduces a novel ZnO@C core-cell structure that significantly improves piezoelectric nanogenerator performances. By synergistically combining ZnO nanofillers with a conductive carbon shell, this design addresses key limitations of traditional ZnO-based PENGs, offering superior charge transfer efficiency, stronger interfacial polarization, and enhanced filler dispersion within PVDF matrix. The carbon coating enhances piezoelectric performance by reducing surface energy, ensuring uniform dispersion, and restricting filler agglomeration. Moreover, the conductive coating facilitates robust interfacial polarization and efficient charge transfer, leading to superior piezoelectric output compared to conventional ZnO-based designs. The high surface-to-volume ratio of ZnO nanorods, combined with the conductive carbon shell, further boosts the electroactive phase content and dielectric properties of the PVDF matrix. These advancements address key challenges in wearable energy-harvesting applications by providing a lightweight, breathable, and water-resistant nanogenerator with improved mechanical integrity and long-term durability. This novel design represents a significant step forward in developing high-performance PENGs for modern energy-harvesting applications.

Thus, this study aims to develop a PENG using ZnO nanofillers in two distinctive morphologies (nanoparticles and nanorods), with a core cell structure coated with carbon (ZnO@C) to enhance piezoelectric performance. The polymeric membrane is fabricated using electrospinning technique to achieve a lightweight, breathable and water-resistant material compared to conventional spin coat or solvent cast method [41, 42]. Additionally, the electrospinning technique enables in situ electrical poling and mechanical stretching, resulting in increased electroactive β-phase formation in PVDF, further enhancing device performance.

## 2. Experimental

### 2.1 Synthesis of Core-cell ZnO@C Nanorods and Nanoparticles

ZnO nanostructures, including both nanorods and nanoparticles, were synthesized using flame spray pyrolysis (FSP) and obtained from the International Advanced Research Centre for Powder Metallurgy and New Materials, Hyderabad, India. Notably, the experiments revealed contrasting results when methanol was substituted for ethanol in the precursor solution, while other process parameters kept constant. In particular, the utilization of methanol as the precursor led to the formation of ZnO nanoparticles, whereas the substitution with ethanol resulted in the production of ZnO nanorods. Detailed findings of this experiment can be found in the work of co-author K. Hembram *et al.* [43, 44].

The next step involved preparing a core-cell structure of the ZnO nanorods and nanoparticles coated with a carbon layer using polydopamine (PDA). The process began with dispersing 1 g of ZnO nano powders in a solution comprising of 68 ml ethanol and 192 ml deionized water, resulting in a total volume of 260 ml. The solution, labeled as Solution A, was ultrasonicated for 30 min to achieve proper dispersion. Subsequently, a separate solution referred as Solution B was prepared by dissolving dopamine hydrochloride in a mixture of 12 ml ethanol and 12 ml deionized water. The two solutions (Solution A and B) were then mixed together and stirred for 14 h using a magnetic stirrer. As a result, dopamine is coated over ZnO nano powders uniformly as depicted in Figure 1. The dopamine-coated ZnO nano powders were washed five times repeatedly with deionized water and subsequently dried in a vacuum oven at 60°C. Finally, to obtain a core-cell structure of the ZnO@C nanorods and nanoparticles, the powders underwent a heat treatment at 550°C for 2 h in a Nitrogen atmosphere, successfully forming the core-cell ZnO@C nanorods and nanoparticles (carbon coated nanoparticle referred as ZnP, nanorod as ZnR). The inset displays a field emission transmission electron microscope (JEOL, 2100F), FE-TEM image of carbon-coated particle, revealing a carbon layer approximately 7.9 nm thick.

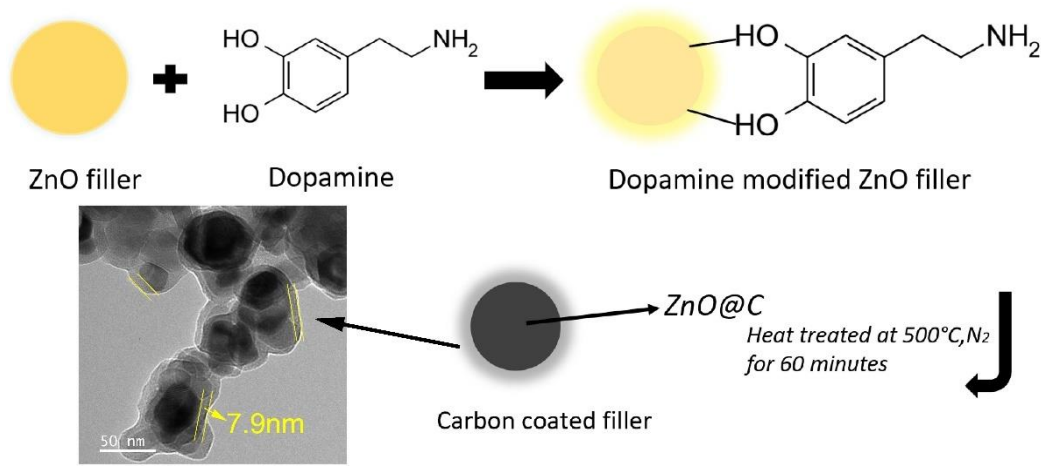

**Figure 1:** A schematic illustrates the formation of core-shell ZnO@C nanostructures using dopamine, with the inset displaying a TEM image of carbon-coated ZnO nanofillers.

**2.2 Preparation and Characterizations of ZnO@C/PVDF Nanofibers**

The ZnO@C/PVDF nanofiber membranes were prepared using a custom-made electrospinning setup (Figure S6), with detailed specifications provided in the Supplementary Data File. For this, ZnO@C nanorods and nanoparticles were used separately in varying weight percentages (0.5, 1.0, and 2.0 wt%,). Initially, PVDF (Sigma Aldrich, with weight-average molecular weight, $M_w$ =1,80,000) was dissolved in a solution of N, N-Dimethylformamide (DMF; Sigma Aldrich) and acetone (Sigma Aldrich) mixed in a 1:1 ratio. Other solution parameters (viscosity, conductivity etc.) were optimized and remain unchanged in this work. This mixture was stirred for 2 h at 60°C to achieve a 20 wt% concentration. ZnO@C nanofillers were then slowly added to the solution as a filler, followed by continuous stirring for 10 h at room temperature to ensure thorough dispersion. The resulting solution was then loaded into a 10 ml syringe for use in the electrospinning process, as illustrated in the schematic of Figure 2. The electrospinning parameters were kept consistent across all samples: a needle-to-collector distance of 12 cm, an 18 kV DC voltage supply, a 22-gauge needle of out diameter of 0.70 mm, and a flow rate of 1 ml/h. A novel spinneret movement design was employed, enabling the needle to move in a to-and-fro motion perpendicular to the plane of the metallic plate collector. This design ensured the large area and uniform thickness deposition of fibres onto aluminium foil [45]. The collected nanofiber membrane/mat dried at 60°C overnight and subsequently utilized in device fabrication. The as prepared membrane was characterized for phase formation using an X-ray diffractometer (Bruker D8 Advance, Germany) with a Cu-K$_\alpha$ wavelength of $\lambda$=1.5404 Å, scanned over a range of 10° to 60° at a rate of 0.02°/min, and a Fourier transform infrared (FTIR) spectroscope (PerkinElmer, USA). Morphological analysis was performed using a scanning electron microscope (SEM) (Tescan MIRA 3 MUG FEG, Czech Republic) was used. The distribution of particle size and fiber diameter was determined from SEM/TEM images using ImageJ 1.44p (National Institutes of Health, USA), a freely available software.

**2.3 Fabrication of ZnO@C/PVDF Energy Harvesting Device**

For the fabrication of the energy harvesting device, we have used aluminium foil (0.025 mm thickness, 99.45% purity, OttoKemi) as electrodes on both sides of the electrospun nanofiber membrane as shown in the Figure 2. To prevent a short circuit between the two opposite electrodes, different dimensions were used, with bottom electrode measuring 2×2 cm$^2$ and top electrode of 1.5×1.5 cm$^2$. The entire device was laminated using a Kapton (polyimide) tape, and copper wires were attached via conductive copper tape (3M, 2245 Conductive Tape, RS India) to ensure proper connectivity, as illustrated in Figure 2. A photo of the fabricated device is also shown in Figure 2. Subsequently, the output voltages from the fabricated device were recorded using a Digital Storage Oscilloscope (DSO) (SDS 1052DL, Siglent Technologies, Netherlands) under varying load resistances (10 kΩ to10$^4$ kΩ) through a rectifier energy harvesting circuit, as illustrated in Figure 2.

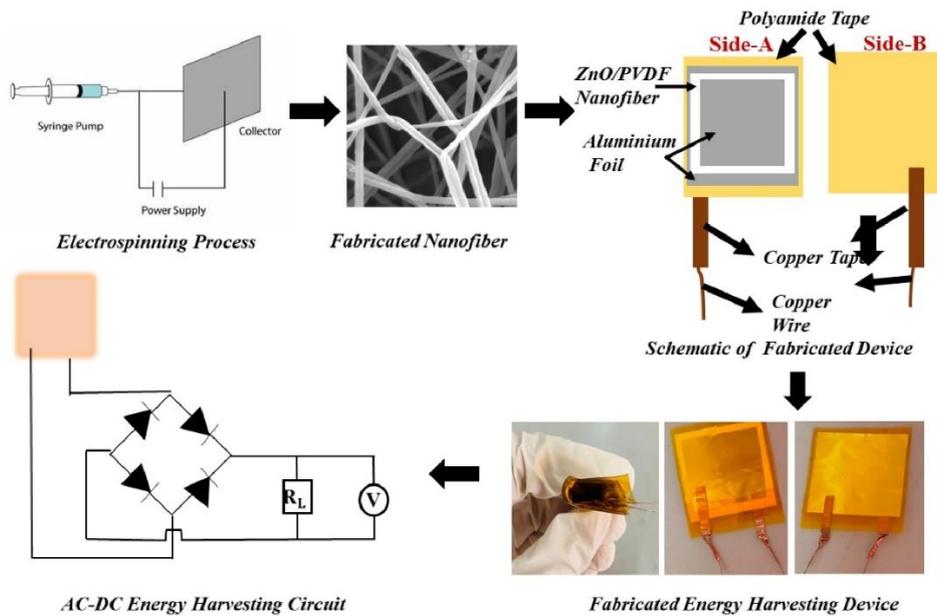

**Figure 2:** Schematic of electrospun nanofiber preparation, flexible energy harvesting device and AC-DC Energy harvesting circuit used to measure piezoelectric output.

## 3. Results and Discussion

### 3.1 ZnO Nanofiller Characterizations

The phase and morphology of ZnO and carbon-coated ZnO nanofillers (ZnO@C) were analysed using X-ray diffractometer (XRD) patterns, scanning electron microscope (SEM) micrographs, and field emission transmission electron microscope (FE-TEM) images, as depicted in Figure 3(a-g). The XRD patterns in Figure 3(g) reveal a hexagonal wurtzite (ZnO) phase for both the fillers, without presence of any secondary phases. All diffraction peaks are indexed to JCPDS 01-079-2205, indicating a highly crystalline structure. The SEM and TEM images, presented in Figure 3(a), (b), (d), and (e), confirm the nanoparticle and nanorod morphologies. The size distribution in Figure 3(h) indicates an average nanoparticles size of 21.9 nm, while Figure 3(i) shows that nanorod of lengths vary from 20 to 80 nm (average nanorod length = 49.9 nm) with diameters ranging from 10 to 20 nm. A uniform thin carbon layer of thickness about 7.9 nm is observed on each ZnO nanofillers, as shown in Figure 3(c) and (f) for nanoparticles and nanorods, respectively.

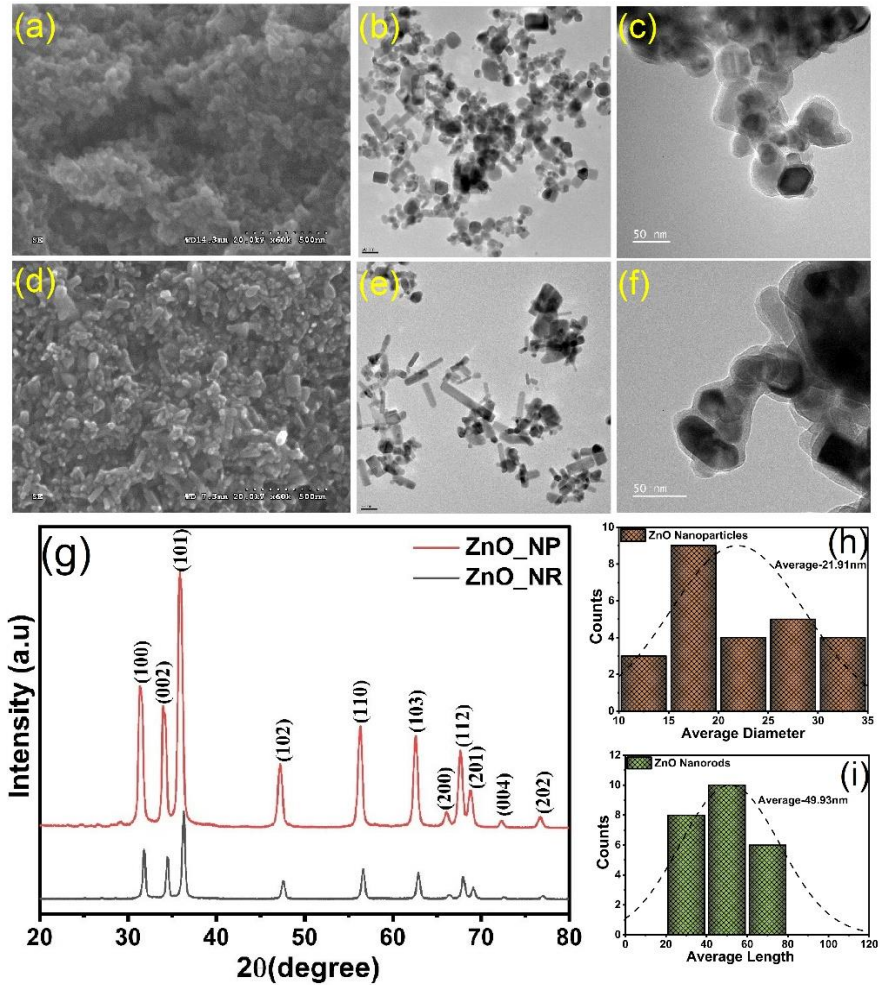

**Figure 3:** ZnO nanofiller characterizations showing (a) SEM, (b) TEM, and (c) TEM of carbon coated nanoparticle; (d) SEM, (e) TEM, and (f) TEM of carbon coated nanorod; (g) XRD patterns of ZnO nanorod and nanoparticle; (h)-(f) particle size distribution of ZnO nanoparticle and rod respectively.

### 3.2 Phase Identification and Quantification

The influence of varying contents of ZnO@C nanofillers (carbon-coated nanoparticles referred to as ZnP, and nanorod as ZnR) on the formation of different phases of PVDF (α, β, γ) was investigated and quantified using X-ray diffraction (XRD) and Fourier-transform infrared (FTIR) spectroscopy. As shown in Figure 4(a), the XRD patterns reveal prominent peaks corresponding to the α-phase (020) at 2θ =18.3°, the β-phase (110)/(200) at 2θ = 20.26°, and a mixed phase of β and γ (110) at 2θ = 20.5°. Additionally, two relatively weak characteristic peaks of the β-phase are observed at 2θ = 36.2º and around 57.0º, corresponding to the (101) and (221) planes, respectively.

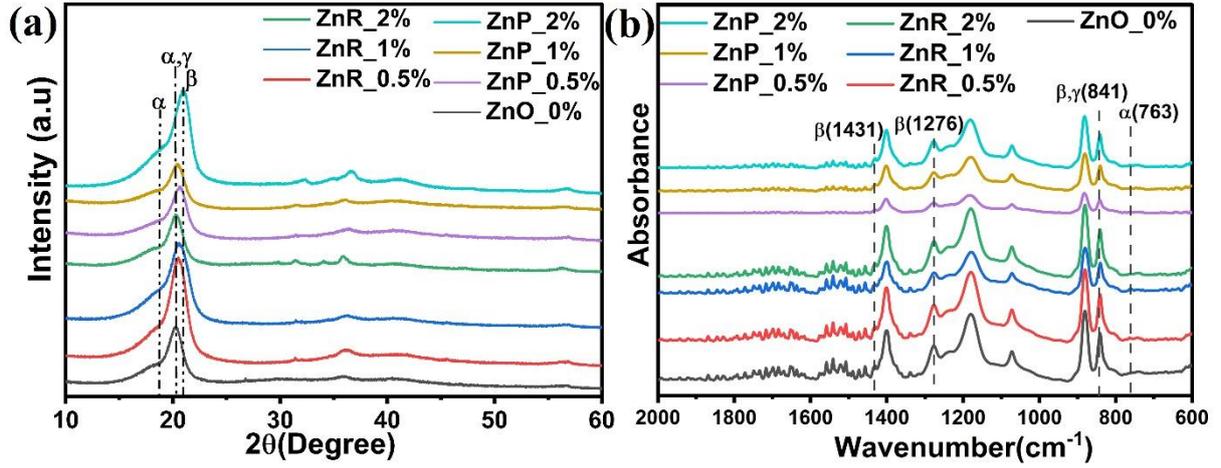

**Figure 4:** (a) XRD patterns and (b) FTIR spectrum of ZnO@C/PVDF nanofibers.

To quantify the individual β and γ phases, which is critical for understanding piezoelectric properties, peak deconvolution was performed. Figure 5 presents the deconvoluted peak profiles in the 2θ range of 10-30° for all the samples. The dominant β-phase peak (110)/(200) at 2θ = 20.26° (in blue) increases with nanofiller content from 0.5% to 2% in both ZnP and ZnR. The α-phase peak (020) at 2θ = 18.3° (in red) is absent in most samples, while appears with very weak intensity in ZnP_1%, ZnP_2%, and ZnR_2%. Another α-phase peak (110) (in light green) is observed in all nanorod samples but is absent in ZnP_1% and ZnP_2% samples. The γ-phase peak (110) is present in all samples except for nanorods with 2% nanofiller (ZnR_2%).

After identification of individual phases (associated α, β, γ-phase peaks), the crystallinity was calculated from equation (1) using data presented in Figure 5. The degree of crystallinity ($\chi_c$) was calculated from,

$$\chi_c = \frac{\sum A_{cr}}{\sum A_{cr} + \sum A_{amr}} \times 100\% \qquad \text{Equation (1)}$$

where, $\sum A_{cr}$ and $\sum A_{amr}$ are the sum of integral area of crystalline peaks and amorphous halo respectively [46]. It is observed that the incorporation of ZnR significantly improves the crystallinity compared to ZnP. The percentage crystallinity ($\chi_c$ %) increases from 41.9 % in pure PVDF to 76.1 % in ZnR_2%, whereas a slight decrease in crystallinity is observed with increasing ZnP filler content, as presented in Table 1 and illustrated in Figure 6(a).

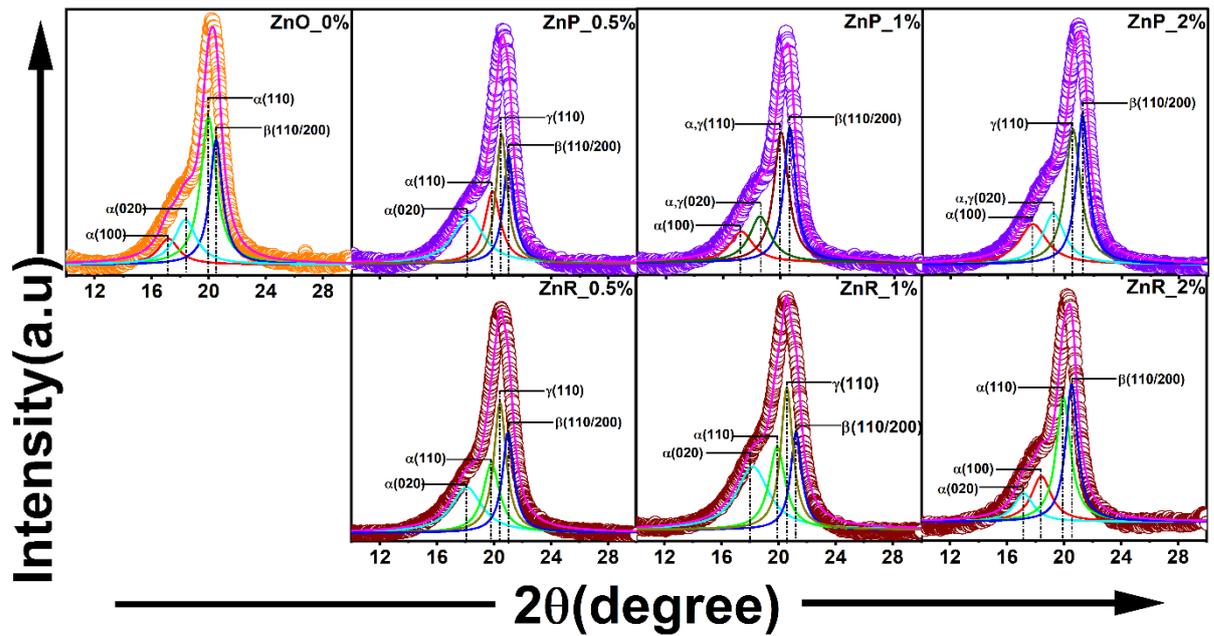

**Figure 5:** Quantification of different phases of XRD for ZnO@C/PVDF nanocomposite film with varying filler content (0.5, 1.0, 2.0 wt%)

FTIR spectroscopy further quantified electroactive phase content, reinforcing the observed trends in the XRD analysis. Figure 4(b) presents the FTIR spectra for all samples. The peak at 763 cm⁻¹ corresponds to the dominant α-phase, while the peak at 841 cm⁻¹ corresponds to contributions from both the β and γ phases of PVDF. To distinguish between these phases accurately, additional characteristic peaks are were referenced: the α-phase is identified by peaks at 763 cm⁻¹ and 614 cm⁻¹, while the β-phase is specifically marked by 1275 cm⁻¹ peak, and the γ-phase is characterized by a distinct peak at 1234 cm⁻¹.

A comprehensive analysis across a broader spectral range confirms that the 841 cm⁻¹ peak predominantly attributable to the β-phase. Figure 4(b) highlights the dominance of the β-phase peak at 841 cm⁻¹, with ZnR_2% exhibiting the most intense β peak, indicating its enhanced electroactive properties. The electroactive phase was quantified utilizing Equation (2), which incorporates the absorbance values corresponding to the dominant alpha (α) peak at 763 cm⁻¹ and the beta (β) peak at 841 cm⁻¹ for all the samples [20]. The respective electroactive β-phase percentages are tabulated and summarized in Table 1 using the equation,

$$F(\beta) = \frac{A(\beta)}{\{K(\beta)/K(\alpha)\}A(\alpha)+A(\beta)} \times 100\% \qquad \text{Equation (2)}$$

where, the electroactive phase content is represented by F(β), and the absorbance at 763 cm⁻¹ and 840 cm⁻¹ are denoted by A(α) and A(β) respectively. For these wavenumbers, the respective absorption coefficients are K(α) and K(β) with values 6.1 × 10⁴ and 7.7 × 10⁴ cm² mol⁻¹, respectively.

**Table 1:** Quantification of overall electroactive (EA) phase (%) for ZnO@C/PVDF nanocomposite film with varying filler content (0.5, 1.0, 2.0 wt%)

| Filler Content (wt%) | $\chi_c$ (%) | | $F_\beta$ (%) | | Overall EA Phase (%) = ($\chi_c \times F_\beta$) | |
|---|---|---|---|---|---|---|
| | (NR) | (NP) | (NR) | (NP) | (NR) | (NP) |
| 0 | 41.9 | 41.9 | 83.0 | 83.0 | 34.8 | 34.8 |
| 0.5 | 68.8 | 68.1 | 93.0 | 91.0 | 64.0 | 61.9 |
| 1 | 72.1 | 64.9 | 92.0 | 92.0 | 66.3 | 59.7 |
| 2 | 76.1 | 65.0 | 92.0 | 95.0 | 69.9 | 61.8 |

The results presented in Table 1 and illustrated in Figure 6 indicate that the electroactive β-phase content increases from 83.0% for pure PVDF to a maximum of 95.0% for ZnP_2% sample, containing 2% ZnO@C nanoparticles. Although the variation in the β-phase content between ZnR and ZnP is insignificant, both morphologies are expected to enhance piezoelectric properties due to their relatively high β-phase proportion (ranging from 91.0 % to 95.0%). However, piezoelectric performance depends on the overall electroactive (EA) phase content, which is determined by the β-phase proportion within the total crystalline phase. In this context, the overall EA-phase is calculated using the respective ($\chi_c \times F_\beta$) % values [46, 47] and is presented in Figure 6(b) and Table 1. It is evident that all the nanorod-incorporated PVDF (ZnR) samples exhibit a higher overall EA-phase content compared to its nanoparticle counterpart (ZnP).

Thus, these results can be attributed to the surface charges and dipole interaction for β-phase crystallization, as well as nanofiller alignment along the fiber axis, as schematically presented in Figure 6(c). Specifically, the carbon layer on the ZnO fillers contains a high density of $sp^2$-hybridized carbon atoms with negative charges [48], which induce an ion-dipole interactions with $CH_2$ dipoles in PVDF. As a result, the interfacial attractive interaction between $CH_2$ ($\delta^+$) groups and the negatively charged carbon surface promotes a folded arrangement of PVDF chains into a lamellar arrangement, favouring β-phase crystallization, as illustrated in Figure 6(c). In pure PVDF, the polar β-phase typically forms due to the elongation and simultaneous orientation of the polymer chains under the electric field during the electrospinning.

Furthermore, the addition of ZnO@C nanofillers into PVDF not only acts as nucleating agent but also significantly enhances β-phase crystallization, particularly in case of nanorods. Because ZnR nanofillers possess a higher surface-to-volume ratio than ZnP (as reported in earlier study by our coauthor Hembram *et al.* [43]), they provide a greater number of nucleating sites, resulting in an increased crystalline phase fraction in ZnR. Additionally, during the electrospinning, ZnR aligns along the fiber axis, inducing lamellae alignment along the fiber axis, thereby further enhancing β-phase formation, as illustrated in Figure 6(c). In contrast, symmetric and isotropic shape of ZnP lacks the anisotropy, resulting in reduced β-phase formation. Similar phenomena have been observed in CNT/PAN nanofibers by Hou *et al.* [49] and theoretically explained by Dror *et al.* for MWCNT-embedded PEO electrospun nanofiber [50]. Overall, these findings highlight the critical role of ZnO@C fillers, particularly nanorods, in enhancing piezoelectric properties through mechanism beyond just the relative electroactive β-phase content.

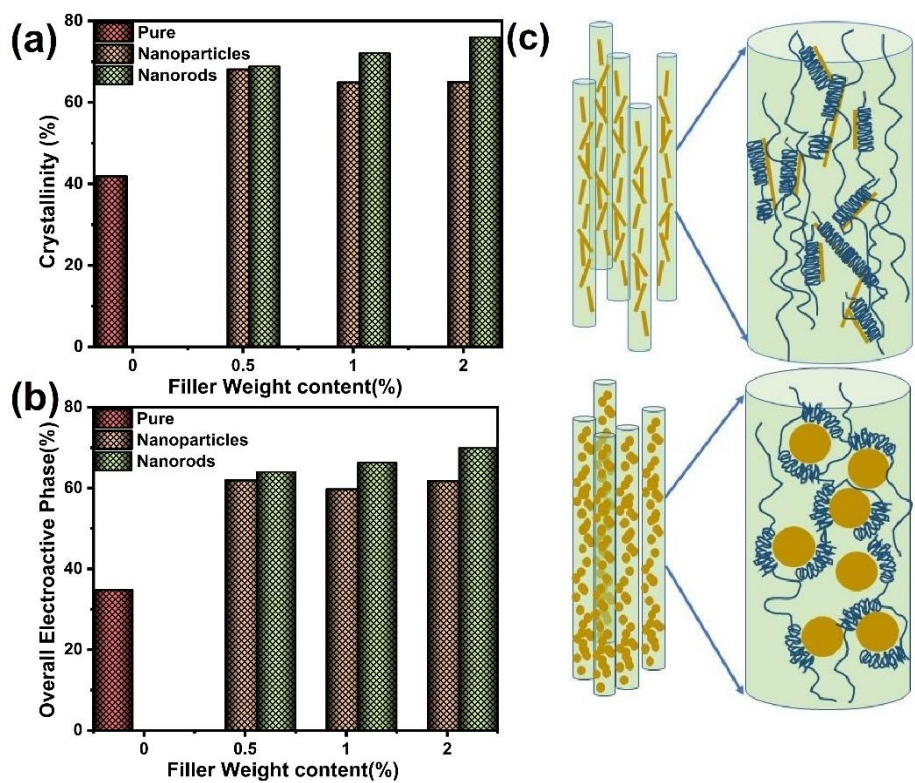

**Figure 6** (a) Crystallinity (%), and (b) overall electroactive phase (%) with varying filler content, (c) schematic illustration of interaction between nanofillers and PVDF polymer chain.

### 3.3 Morphology Study

The SEM micrographs of ZnO@C/PVDF nanofibers and the corresponding fiber diameter distribution of the membranes are illustrated in Figure 7, demonstrating the structural stability of the incorporated nanoparticles and nanorods within PVDF fibers. The adjacent magnified images reveal that incorporating these fillers into the PVDF nanofibers enhances the alignment and distribution of the nanoparticles and nanorods within the fibers, leading to increased stiffness and overall structural reinforcement. The diameters of the nanofiber were found to range from 161 nm to 314 nm. An increase in the concentration of ZnR and ZnP in PVDF during electrospinning not only results in larger nanofiber diameters but also influences the electroactive phases of PVDF. The addition of ZnO filler alters the flow behaviour of the solution, which impedes the stretching and elongation of the polymer jet, resulting in the formation of thicker fibers. Therefore, the incorporation of ZnO not only influences the fiber diameter through viscosity changes but also significantly affects the electroactive phase content, enhancing the functional properties of PVDF nanofibers.

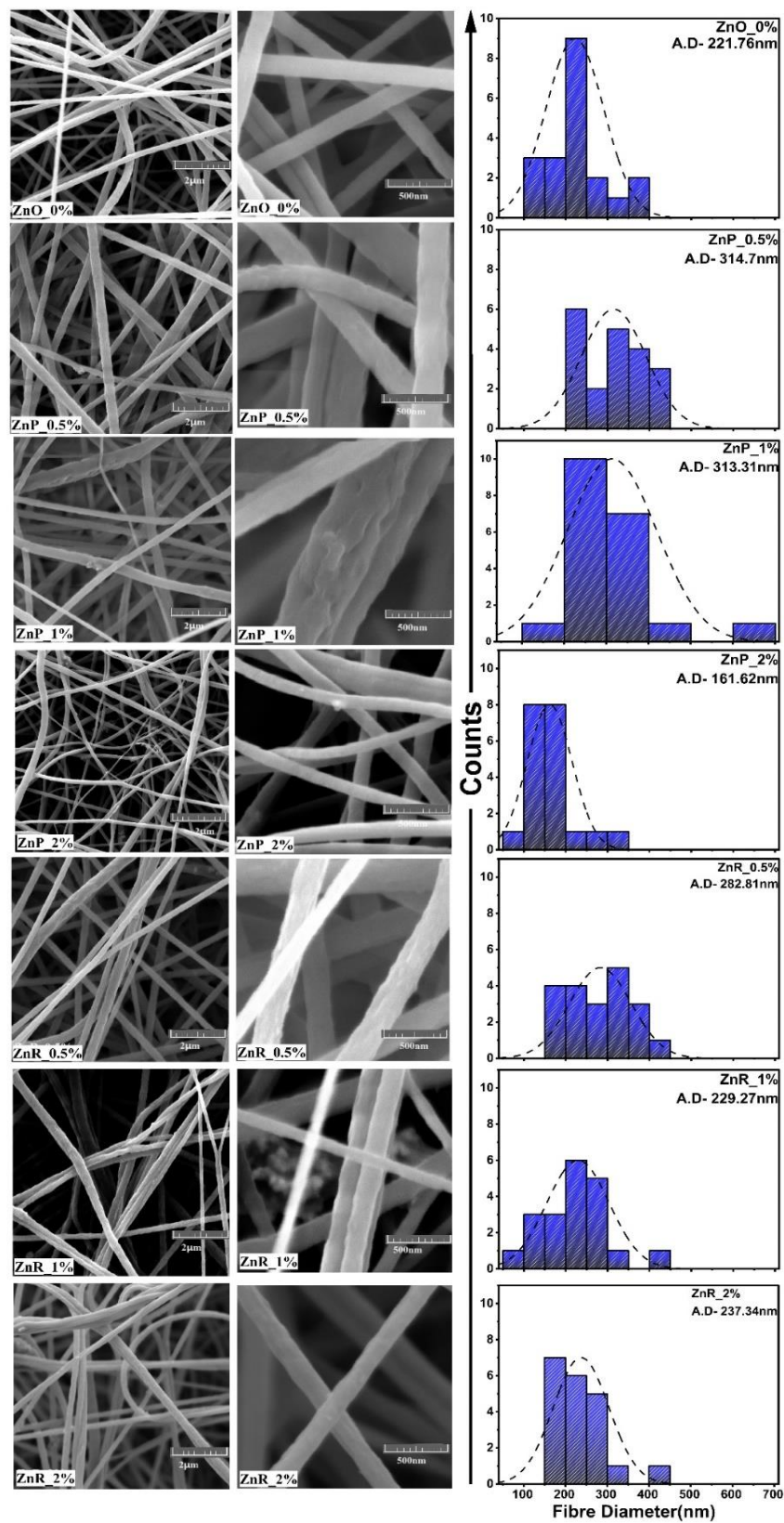

**Figure 7:** SEM micrographs and distribution of fiber diameter of all ZnO@C/PVDF nanofibers.

### 3.4 Breathability and Waterproof Performance

The nanofiber membranes utilized for energy harvesting in wearable applications are highly sought after for their breathability, which ensures comfortable skin contact, and their waterproof nature. The breathability of the membrane was evaluated by measuring the water vapour transmission rate (*WVTR*) using Fickian diffusion relation, following ASTM E 96-97,

$$WVTR = \frac{M}{A.t} \qquad \text{Equation (3)}$$

where, *M* is the weight loss of water (in kg), *A* is the test surface area (in m$^2$), *t* is the time duration (days, with 1 day = 24 h).

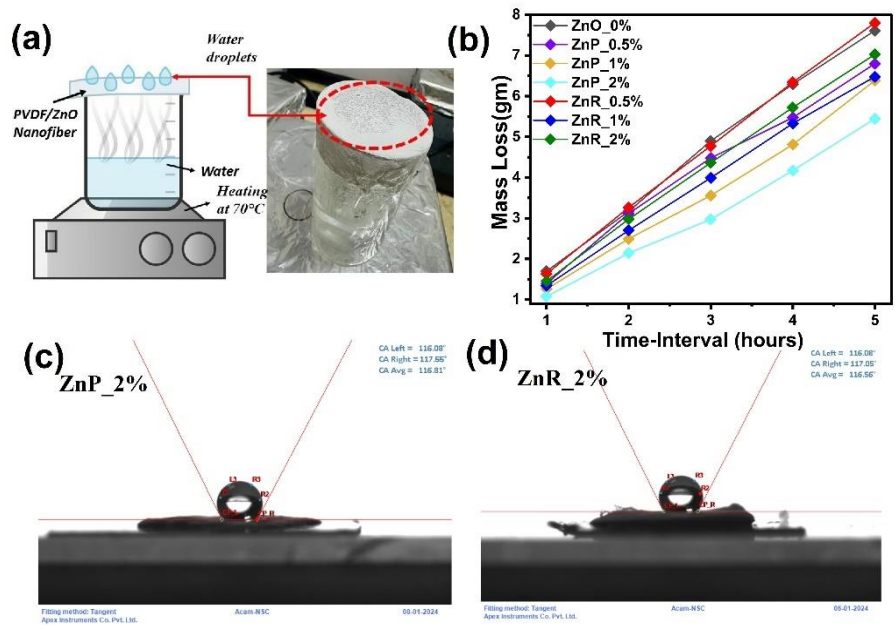

**Figure 8:** (a) Schematic and process of mass loss from beaker through nanofiber membrane (b) mass-loss per hour in nanofiber membrane (c-d) water contact angle measurement of nanofiber membrane for measuring its hydrophobicity.

A schematic of the experimental setup is shown in Figure 8(a). A beaker containing water was fully covered by the nanofiber membrane on its upper surface and placed on a hotplate set to a temperature of 70°C. As the water began to boil, vapour passed through the nanofiber membrane at the top of the beaker. Another beaker, positioned above the first beaker, revealed visible condensation of water droplets on the upper side of the nanofiber, and the presence of vapours was noticeable on the beaker placed above, as illustrated in Figure 8(a). This observation clearly demonstrates substantial water droplet condensation on the upper side of fibre, providing convincing evidence for the notable water vapour permeability, confirming the breathability of the PVDF/ZnO nanofiber membrane. However, the breathability was quantitively evaluated using WVTR by measuring the mass loss of water over a specific time duration, as shown in Figure 8(b). The WVTR values range from 0.4 to 0.6 $kg.m^{-2}.d^{-1}$ for all samples, confirming water vapor permeability comparable to other fiber-based nanogenerators, including cotton woven fabric [51].

The waterproof characteristics of the membrane were assessed through water contact angle (WCA) measurements utilizing the sessile drop method. A water droplet was carefully positioned above each fiber membrane, and subsequent WCA measurements were conducted employing a contact angle measuring microscope [34]. All fiber membranes exhibited notable hydrophobicity, However, as representative compositions, the WCA of samples with ZnO@C nanorods (ZnR_2%) and nanoparticles (ZnP_2%), which showed contact angles of 116.56° and 116.81°, respectively, is illustrated in Figure 8(c-d). These results affirm the inherent waterproof nature of the nanofibers.

### 3.5 Dielectric and Piezoelectric Measurements

The frequency-dependent dielectric properties reflect different polarization mechanism within a dielectric material under an alternating electric field, as well as charge transfer effects that contribute to improved piezoelectric performance. Figure 9 illustrates the frequency-dependent dielectric constant and loss, along with the variation of dielectric constant for different filler compositions of ZnO@C/PVDF samples. As seen in Figure 9(a), the dielectric constant remains nearly constant up to ~ $10^4$ Hz for all samples and then rapidly decreases with increasing frequency. This is due to accumulation of charges at the heterogeneous dielectric interfaces between carbon-coated ZnO and PVDF, which eventually fail to follow the applied field at higher frequencies (> $10^4$ Hz). This higher value of the dielectric constant is attributed to strong interfacial polarization (Maxwell-Wagner-Sillars, MWS), which enhances dipole alignment within the PVDF matrix. At low frequencies, dielectric loss primarily arises from interfacial polarization and space charge accumulation, whereas at high frequencies, it is attributed to charge mobility within the conductive carbon shell, which increases leakage currents. However, the frequency-independent and moderate dielectric loss value ranging from 0.4 to 0.5 across all ZnO@C/PVDF nanofiber membranes suggests efficient charge accumulation with minimal leakage.

Furthermore, the overall dielectric constant increases consistently with increasing ZnR filler content. In contrast, for ZnP samples, the dielectric constant initially increases up to 0.5% filler concentration but then declines. This trend can be explained by the high surface-to-volume ratio of ZnR, which offers larger interfacial area for interaction with PVDF, thereby enhancing dipole alignment and interfacial polarization. At higher ZnP content, excessive filler aggregation leads the system to a higher percolation threshold, disrupting the insulating PVDF matrix and reducing the dielectric constant.

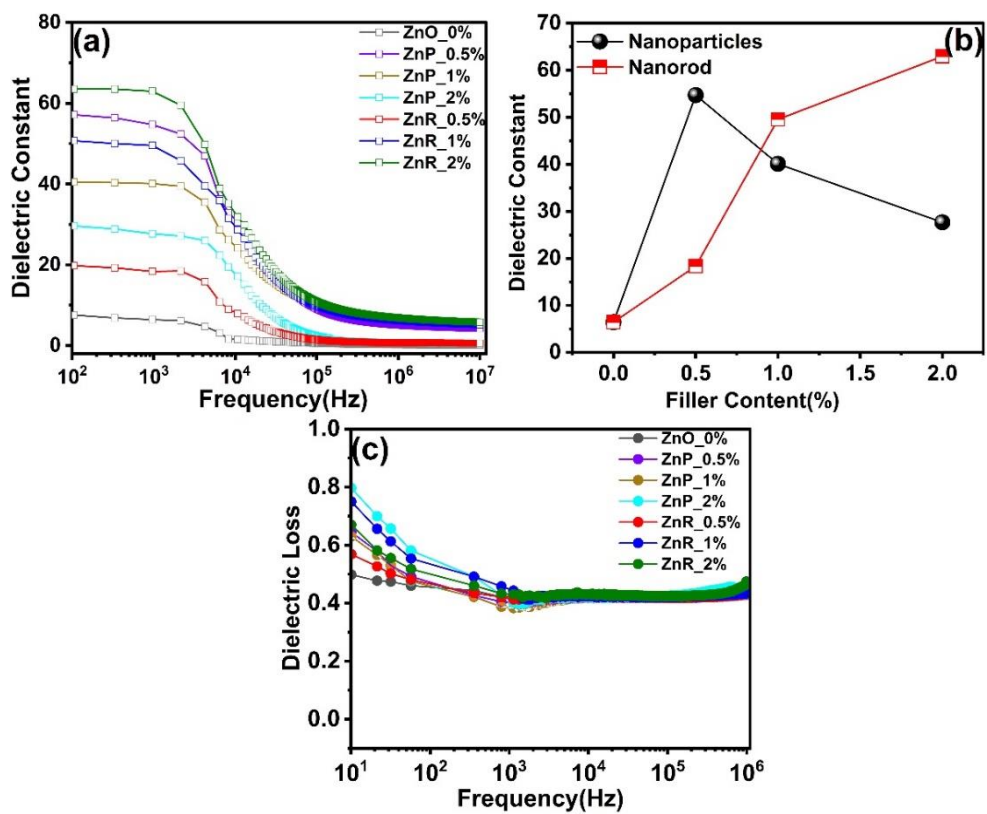

**Figure 9:** Variation of dielectric constant with (a) frequency, (b) different filler content, and (c) of dielectric loss with frequency for all the ZnO@C/PVDF samples.

The energy harvesting data were acquired using a digital storage oscilloscope (DSO) through continuous human finger tapping on the device surface. Under a periodic finger imparted force (~ 25 N, frequency = 4 Hz, refer to Supplementary Data File for detailed calculations), the piezoelectric effect facilitated the accumulation of electric charges, converting mechanical energy into electrical energy. Typically, the piezoelectric output is recorded as open-circuit voltage ($V_{oc}$) and short-circuit current ($I_{sc}$), with their corresponding power density calculated as, $V_{oc} \times I_{sc}$. However, from the practical aspect, a nanogenerator must be connected to a load resistance ($R_L$) rather than operating under open- or short-circuit conditions to avoid overestimating power density [52]. Therefore, the rectified output voltage ($V_{out}$), obtained using a rectifier circuit (Figure 2), was carefully measured under varying load resistance ($R_L$) for each sample, as illustrated in Figure 10(a). The corresponding power density of the device across an externally connected variable load resistance ($R_L$) ranging from 10 kΩ to $10^5$ kΩ was measured, as shown in Figure 10(b). The instantaneous power was calculated using the equation, $P = V_{out}^2/R_L$, and power density was determined by using $P/$ (area × thickness) and is listed in Table 2.

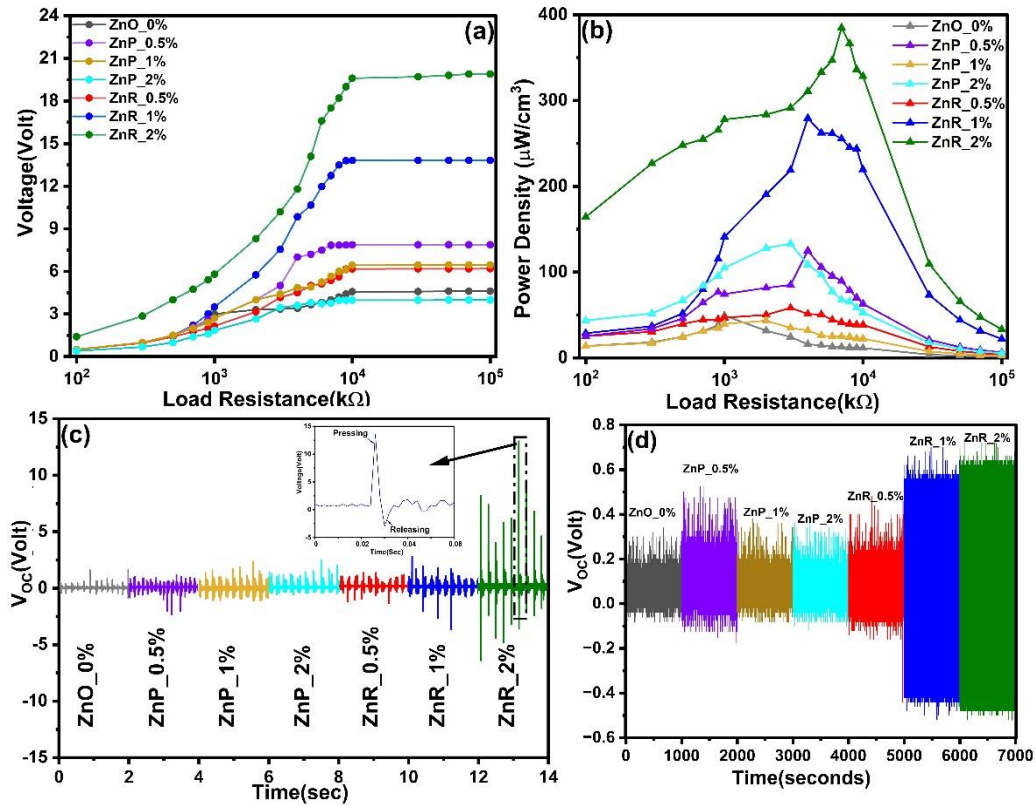

**Figure 10:** (a) Piezoelectric output voltage vs load resistance (b) power density vs load resistance (c) Open-circuit voltage ($V_{oc}$) vs time (d) cyclic stability test for all the ZnO@C/PVDF devices.

**Table 2:** Piezoelectric output voltage and power density of ZnO@C/PVDF devices.

| Sample Id | $V_{OUT}$ (Max) (Volt) | Maximum Power Density ($\mu W/cm^3$) | Load Resistance ($R_L$) (M$\Omega$) @$P_{max}$ |
|---|---|---|---|
| ZnO_0% | 4.6 | 49.26 | 1 |
| ZnP_0.5% | 7.86 | 124.98 | 4 |
| ZnP_1% | 6.45 | 43.40 | 2 |
| ZnP_2% | 3.98 | 133.17 | 3 |
| ZnR_0.5% | 6.19 | 58.33 | 3 |
| ZnR_1% | 13.83 | 279.44 | 4 |
| ZnR_2% | 19.9 | 384.83 | 6 |

A maximum power density of 384.83 µW/cm$^3$ was obtained at $R_L = 10^4$ k$\Omega$, with a corresponding maximum $V_{out}$ = 19.9 V for the ZnR_2% samples. In the study involving nanoparticles (ZnP), the highest output voltage of 4.24 V was recorded at a filler concentration of 0.5%, resulting in a maximum power density of 124.98 µW/cm³ for ZnP_0.5%. For nanorods, an output voltage 19.9 V and maximum power density 384.83µW/cm$^3$ were observed at 2% filler concentration (ZnR_2%).

Furthermore, to evaluate a more realistic output performance of the piezoelectric membrane, open-circuit voltage ($V_{oc}$) vs. time was recorded at load resistance, $R_L$ =10$^4$ $\Omega$, that corresponds to the value at maximum power density (Figure 10(b)). The results are presented in Figure 10(c). The time-resolved peak of $V_{oc}$ shown in the inset during a pressing and releasing confirms the output is due to piezoelectric effect. The long-term electromechanical stability was evaluated through a custom-made mechanical stimulation setup (Supplementary Data File, S7) recorded over 1000 seconds for each sample. The results, shown in Figure 10(d), indicate consistent output from the developed PENG, confirming its suitability for practical applications.

Despite minimal variation in the EA-phase between ZnP and ZnR, the piezoelectric output differs significantly due to differences in charge mobility and interfacial polarization. This confirms that the carbon shell is the key for obtaining superior performance of ZnO@C-based PENGs, regardless of their morphology or filler content. The piezoelectric potential generated within the nanofillers under mechanical force induces the movement and accumulation of mobile charges on the carbon-layer, amplifying the voltage and enhancing the piezoelectric output. Additionally, carbon-layer facilitates strong interfacial polarization between nanofillers and polymer matrix, as confirmed from dielectric measurements.

Furthermore, Table 3 represents a comparison of different synthesis methods and their corresponding output with the finding of this study. Compared to previous investigations on PVDF/ZnO nanocomposite films, the current study has achieved a substantially higher open circuit voltage ($V_{oc}$ ~ 19.9V) [32, 37, 53-57].

**Table 3:** Comparative study of different method and their corresponding piezoelectric output with present study.

| Materials Used | Synthesis Method | Power Density (mW/cm²) | Output Voltage (Volts) | Mode of Testing | Ref |
|---|---|---|---|---|---|
| PVDF/ZnO nanocomposite | Spin coating | 0.22 | 5.3 | Bending motion | [32] |
| CNT/ZnO/PVDF | Solution Casting | 0.66 | 1.32 | Mechanical Deformation | [58] |
| ZnO microrods/PVDF | Supersonic spraying | 12.5 | 15.2 | Tapping | [59] |
| Breathable ZnO@PVDF fibrous nanogenerator | Electrospinning | 0.52 | 12.6 | Wearable applications | [30]. |
| ZnO/PVDF composite | Spin Coating | - | 4.2 | Hand-Pressing | [60] |
| Paper ash-ZnO/PVDF nanofibers | Electrospinning | 1.1 | 8.3 | Wind energy harvesting | [31] |
| PVDF/ZnO@C nanocomposite fibers | Electrospinning | 45.92 | 19.9 | Hand Tapping | **This Study** |

## 4. Conclusions

In conclusion, carbon-coated core cell ZnO nanostructured fillers (ZnO@C) with nanoparticle (ZnPs) and nanorod (ZnRs) morphologies were successfully incorporated into PVDF nanocomposites, significantly enhancing their piezoelectric performances. Electrospinning not only facilitated the formation of the electroactive β-phase in PVDF but also imparted breathability and waterproof properties. The enhanced piezoelectric response is primarily attributed to the carbon coating on ZnO, which effectively enhances the ion-dipole interaction between the negatively charged carbon and the $CH_2$ groups of the PVDF chain. These interactions promote a well-ordered arrangement of PVDF chain into an electroactive β-phase. Notably, while both ZnPs and ZnRs exhibited minimal variation in the electroactive phase, ZnRs demonstrated superior voltage output. This highlights the pivotal role of conductive carbon layer in achieving enhanced performance in ZnO@C based PENGs, irrespective of their morphology or filler content. Under mechanical stress, the nanofillers undergo self-polarization through the migration and accumulation of mobile charges on carbon layer, essentially amplifying the overall voltage output. Furthermore, the superior piezoelectric performances of nanorods can be attributed to their high anisotropy and larger surface-to-volume ratio compared to nanoparticles. The conductive carbon layer on the nanofillers enhances interfacial polarization, further improving piezoelectric performance, while the reduced surface energy of ZnO@C ensures better dispersibility within the PVDF matrix.

A breathability test, measured using the water vapour transmission rate (WVTR), yielded a value of ~ 0.5 $kg/m^2/day$, indicating excellent breathability. Additionally, the waterproofness, calculated through water contact angle measurements, showed a value of ~116°, confirming strong waterproof properties. The innovative idea of introducing a thin conducting layer between nanofillers and polymer matrix offers an active area for mobile charge to accumulate and polarize the nanofillers, significantly boosting the piezoelectric output. These findings establish the ZnO@C/PVDF membrane as an efficient wearable piezoelectric nanogenerator, paving the way for self-powered electronics, biomedical sensor, and flexible energy harvesting systems.

**Acknowledgement:** Not Applicable

**Conflict of interest:** All authors declare that there is no conflict of interest.

**Author Contribution (CRediT):**

Anshika Bagla: Performed experiment, Writing- original draft, Data curation, Formal Analysis, Investigation

Kaliyan Hembram: Provided ZnO powders, Validation, Investigation

François Rault: Writing- review & editing, Validation

Fabien Salaün: Writing- review & editing, Supervision

Subramanian Sundarrajan: Writing- review & editing

Seeram Ramakrishna: Writing- review & editing, Supervision

Supratim Mitra: Conceptualization, Methodology, Supervision, Writing- original draft, Validation

**Data Availability:** Not Applicable

**Supplementary Data File:** Available

**Ethical approval:** Not Applicable

**TOC**

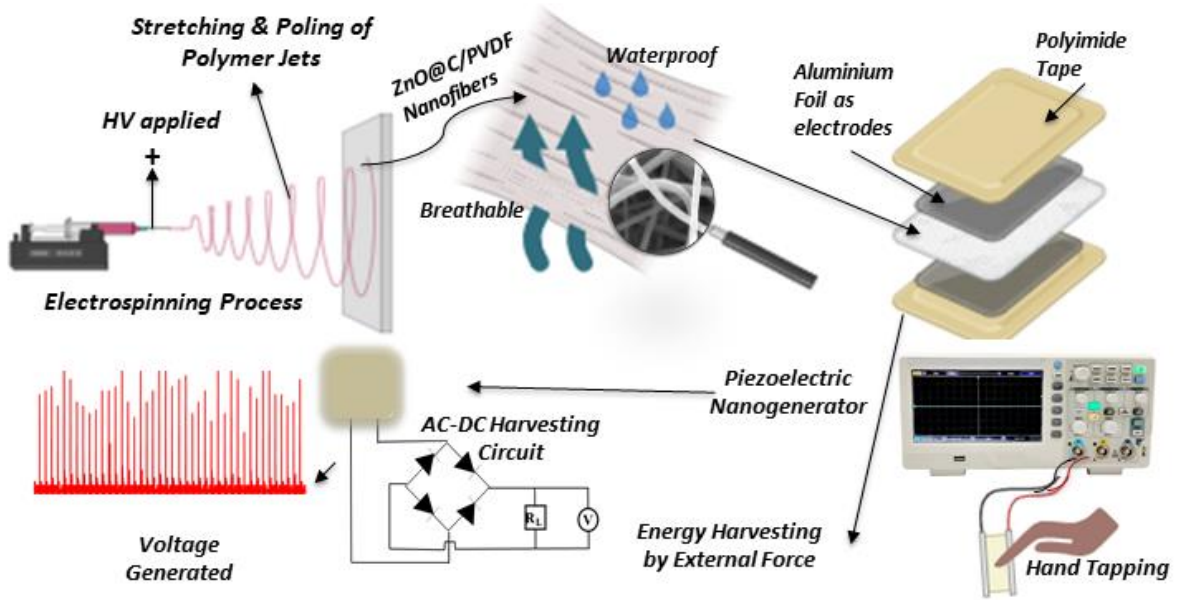